\documentclass{article}

\usepackage{amsmath}
\usepackage{amsfonts}
\usepackage{amssymb}

\usepackage{graphicx}

\begin{document}

\title{Fast Autocorrelated Context Models for Data Compression}

\author{John Scoville}
\maketitle

\begin{abstract}
A method is presented to automatically generate context models of data by calculating the data's autocorrelation function.  The largest values of the autocorrelation function occur at the offsets or lags in the bitstream which tend to be the most highly correlated to any particular location.  These offsets are ideal for use in predictive coding, such as predictive partial match (PPM) or context-mixing algorithms for data compression, making such algorithms more efficient and more general by reducing or eliminating the need for ad-hoc models based on particular types of data.  Instead of using the definition of the autocorrelation function, which considers the pairwise correlations of data requiring $O(n^2)$ time, the Weiner-Khinchin theorem is applied, quickly obtaining the autocorrelation as the inverse Fast Fourier Transform of the data's power spectrum in $O(n\log n)$ time, making the technique practical for the compression of large data objects.  The method is shown to produce the highest levels of performance obtained to date on a lossless image compression benchmark.
\end{abstract}

\section{Introduction}
\verb" "
\indent
Instead of being specified \emph{a priori} in a customized model for data compression, regions of related data ("contexts") may be determined automatically using statistical correlations.

Conventionally, correlation is a pairwise process having a time complexity of $O(n^2)$, which makes a full-scale calculation of correlations impractical for arbitrarily large data. However, for the purposes of data compression, we wish to determine the offsets or lags which tend to be the most highly correlated with a current location in the bitstream.  This is given by the offsets having the largest values of the \emph{autocorrelation} function\cite{VC92}, whose value is defined through the correlation of data with a lagged or shifted version of itself.  If calculated directly from the definition, this is an $O(n^2)$ process.  However, the autocorrelation may be calculated using the Weiner-Khinchin theorem\cite{VC92}, which states that the autocorrelation is the inverse Fourier transform of the power spectrum.  Since the power spectrum can be estimated with a Fast Fourier Transform of time complexity $O(n \log n)$ followed by a vector product which is $O(n)$, and subsequently transformed into an autocorrelation using an inverse Fast Fourier Transform which is also $O(n \log n)$, the overall time complexity is $O(n \log n)$.  This makes the method scalable to large data objects, and hence appropriate to use for general data compression.

For example, context-mixing data compression algorithms\cite{MM00} have, in recent years, produced the most compact known representations of general data.  In order to operate efficiently, however, some prior knowledge of the data is necessary, in order to identify regions of data ('contexts') which share information, rather than simply relying on adjacent positions in a bitstream, as is the case with, e.g. dictionary-type archivers.  As a result, implementations of such algorithms, such as the PAQ series of archivers, tend to include a number of customized models which apply to different types of files.  For example, specialized models for image files typically select neighboring pixels as contexts.  For a typical bitmap, neighboring pixels to the right or left of the current position are adjacent to the current position in the stream of underlying data, but pixels above and below occur at offsets (relative to the current position in the bitstream) differing by a factor proportional to the width of the image.  Prior knowledge of the format and parameters (width, number of channels, bit depth, etc.) of an image file allows a customized model to use non-local contexts in the data to obtain much higher levels of predictive compression performance.  The method under consideration presents a very general alternative to filetype-specific context models.

\section{The Autocorrelation of Data}
\verb" "
\indent
The autocorrelation function measures the correlation of a system with itself at different points in time\cite{VC92}.  Its definiton follows from the standard statistical definiton of a correlation coefficient.  The autocorrelation is defined as:

\begin{equation}
R(s,t) = \frac{E[(X_t - \mu_t)(X_s - \mu_s)]}{\sigma_t\sigma_s}
\end{equation}

Where $\mu_t$ and $\sigma_t$ are the mean and standard deviation at time t, respectively, and E is the expectation value (averaging) operator.  It is more commonly expressed under an assumption of stationarity, meaning that $\mu$ and $\sigma$ do not change with time. Then the autocorrelation is a function of the difference between times s and t, the time lag $\tau = s-t$:

\begin{equation}
R(\tau) = \frac{E[(X_t - \mu)(X_{t+\tau} - \mu)]}{\sigma^2}
\end{equation}

We will regard a string of binary data as a system measured by random variable $X$, and the position in that string as being analogous to a time parameter $t$.  The time lag $\tau$, then, represents a context in the data, an offset from the current position in the bitstream.  The values of $\tau$ which maximize this function are the most highly correlated contexts, and we wish to select these offsets for use in a context model.

The above definitions effectively convert signal values into Z-scores, which cast measurement values as the number of standard deviations they are above or below the mean.  This implicitly assumes a Gaussian distribution, and often this normalizing step can be omitted, using the actual un-normalized values of the random variable $X$.  In such a case, the autocorrelation of a real variable $X$ is defined more simply in terms of a vector inner product:
\begin{equation}
R(X,\tau)=\sum_t X_t X_{t+\tau}
\end{equation}

Since we are interested only in the lags having the largest values of the autocorrelation function, rather than the particular maximal values, we refer to this form as the autocorrelation of the data.  The set of lags $\tau$ resulting in the $n$ largest values are the most autocorrelated contexts of the data, we will refer to this set as $A_n(X)$.  If the $(n+1)$th largest value of $R(X,\tau)$ is used as a cutoff threshold, c, the set of $n$ autocorrelated contexts is:
\begin{equation}
A_n(X)=\{\tau:R(X,\tau)>c\}, |A_n(X)|=n
\end{equation}

Unless there is a reason to do otherwise (for example, if we know that the data contains 16-bit samples) calculate the autocorrelation function (and associated Fast Fourier Transforms) as a sequence of 8-bit bytes.

\section{The Power Spectrum of Data}
\verb" "
\indent
The above definition, as noted in the introduction, has time complexity $O(n^2)$ which would make context modeling of large data scale poorly.  The Weiner-Khinchin theorem allows the time complexity to be improved to $O(n \log n)$.  To apply this theorem, we must first estimate the power spectrum of a process which produced the data.

The power spectrum is the amount of energy per unit of frequency in the data\cite{VC92}.  It could be expressed, for example, in watts per Hz.  This is analogous to a frequency spectrum of light or sound, the only difference being that the signal we associate with general data could represent any type of data, executable code, for instance.  One way to estimate the power spectrum is using a discrete Fourier transform, such as the Fast Fourier Transform, and this has the advantage of running in the desired $O(n \log n)$ time.

The Fourier transform may be defined for continuous variables as:
\begin{equation}
FFT(f)=\frac{1}{\sqrt{2\pi}}\int_{-\infty}^\infty f(t)e^{-i\omega t}\,dt
\end{equation}

Where the normalizing constant $\frac{1}{\sqrt{2\pi}}$ is a matter of convention.  The Discrete Fourier Transform simply replaces the integral with a sum:
\begin{equation}
FFT(f)=\frac{1}{\sqrt{2\pi}}\sum_{t=-\infty}^\infty f(t)e^{-i\omega t}
\end{equation}

The Fourier transform results in a complex-valued function having amplitude and phase information about each frequency present in the data.  The power spectrum is simply the square of the amplitudes.  The operation of complex conjugation allows us square the amplitudes while discarding their associated phase information.

The complex conjugate $\overline{z}$ of a complex variable $z=a+ib$ is $\overline{z}=a-ib$, where $i$ is the unit imaginary $\sqrt{-1}$.  For a complex matrix or vector $Z$, it is customary to define the Hermitian conjugate $Z^*$, which conjugates each element of the transposed matrix, $Z^*=\overline{Z^T}=\overline{Z}^T$.  If $Z_{ij}=z$, then $Z^*_{ji}=\overline{z}$.  With this notation, the power spectral density may be written simply as:

\begin{equation}
S(X)=(FFT(X))^* FFT(X)= \overline{FFT(X)} \cdot FFT(X)
\end{equation}

Where $FFT(X)$ is the Fast Fourier Transform of the data X (or the Discrete Fourier Transform calculated by any other means) as a column vector.

It should be emphasized that since the process is presumed to be sampled along a set of discrete points, this is an \emph{estimate} of the Power Spectrum.  The Power Spectrum could be estimated, for example, via the Maximum Entropy method\cite{Jaynes}.  Not all signals have a well-defined Fourier transform, particularly if Plancharel's theorem does not apply, but a spectrum may be obtained by Fourier transforming an autocorrelation obtained directly from its definition.  This notion is based on the Weiner-Khinchin theorem, which we will discuss next.

\section{Applying the Weiner-Khinchin Theorem}
\verb" "
\indent
The Weiner-Khinchin theorem\cite{VC92} states that the autocorrelation function $R(X,\tau)$ is the inverse Fourier transform of the power spectral density $S(X,k)$.  The inverse Fourier Tranform is, of course, the inverse of the Fourier Transform.  For continuous variables:
\begin{equation}
FFT^{-1}(f)=\sqrt{2\pi}\int_{-\infty}^\infty f(t)e^{-i\omega t}\,dt
\end{equation}

By substituting the definition of $FFT(f)$ for $f$ in the above expression, it may be seen that this is the inverse, as $FFT^{-1} FFT(f)=f$.  Again, the normalizing constant $\sqrt{2\pi}$ is a matter of convention, and the discrete case simply replaces the integral with a sum:
\begin{equation}
FFT^{-1}(f)=\sqrt{2\pi}\sum_{t=-\infty}^\infty f(t)e^{-i\omega t}
\end{equation}

With this notation, the Weiner-Khinchin theorem may be expressed as:
\begin{equation}
R(X,\tau) = FFT^{-1}(S(X,k))
\end{equation}

If we estimate S(X,k) via the FFT, we have, explicitly,
\begin{equation}
R(X,\tau) = FFT^{-1}((FFT(X))^* FFT(X))
\end{equation}

Which has the desired $O(n \log n)$ time complexity.  The values of the lag $\tau$ where $R(X,\tau)$ obtains its $n$ largest values are the elements of the set of autocorrelated contexts, $A_n(X)$.  Having calculated this set efficiently, the resulting context model may be applied to the compression of arbitrary data.  Using a model with more highly correlated input data tends to improve the reliability of estimates taken from their associated conditional probability distributions, which leads in turn to superior performance in predictive coding schemes.

\section{Context-Mixing using Autocorrelated Contexts}
\verb" "
\indent
We will apply the context model to general-purpose data compression using context-mixing neural nets.  The set of autocorrelated contexts could be applied in many other predictive coding schemes, for example, the set could form inputs for the Markov chains used by prediction by partial match (PPM)\cite{BWC89} compressors.  The autocorrelated context set efficiently provides highly correlated inputs, which naturally tend to yield better prediction than less correlated inputs which are selected simply based on adjacency in the data stream.  Context-mixing\cite{MM00} is a convenient example, however, and represents the state of the art in compression performance, as exemplified by the PAQ family of compressors.  We will compare a context model generated from the autocorrelation function with the latest of the PAQ series of compressors, PAQ8pdx v4, using both the built-in image model and the default model it uses for general data when a customized model is unavailable.  Additionlly, we will improve upon its built-in image model using the contexts produced from the set of optimally autocorrelated contexts, $A_n$, breaking a previous record for image file compression.

The technique was applied to image compression using a variant of the PAQ algorithm, which employs a context-mixing neural network\cite{MM00}.  Its standard image context model (which combines the values of neighboring pixels in the image using a single neural net) was supplemented with the set $A_n$ of autocorrelated contexts.

The actual contexts supplied to the context mixer, which is identical to the mixer used in PAQ8, are functions of the set of optimal contexts.  The method does not consider the type of data (in this case, an image) supplied.  These contexts combine the optimal set $A_n$ in several ways which we will describe briefly using pseudo-code.

Let \verb"buffer[i]" represents the data at position \verb"i" in an array of bytes, and let \verb"offsets[k]" represent the k'th optimal context, ordered in an array of integers by descending significance.  Let \verb"p" represent the current position in the bitstream.  Let \verb"set(value)" represent the function adding a context, \verb"value", to the mixer, where contexts with multiple inputs are combined using a hash function.  In PAQ8, this function is \verb"cm.set(cx)", where the value cx is a 32-bit integer.  A hash function \verb"hash(n1,n2,..)" is used to combine multiple integer-valued inputs into a single 32-bit integer value for use as a context, as shown in the examples below.

The first type of context is simply the value of the data at the lagged offsets, with a second argument, color, to indicate which color channel (e.g. Red, Green, Blue) applies to position p.  Application of the method to grayscale images proceeds similarly, but in this case the color argument may be omitted as it does not provide useful information in the case of a single channel.  The contexts generated from the set $A_n$ of autocorrelated contexts are, in C-style syntax:\\
\verb"set(hash(buffer[p-offsets[k]],color));"\\
\indent
The second type of context takes the difference of the value of the data at position p and the value of the data at the lagged offsets:\\
\verb"set(hash(buffer[p]-buffer[p-offsets[k]],color));"\\
\indent
The third type of context sums the value of the data at position p with the value of the data at the lagged offsets:\\
\verb"set(hash(buffer[p]-buffer[p-offsets[k]],color));"\\
\indent
The sum operation is orthogonal to the difference operation, which improves coverage and hence compression performance.

The fourth type of context combines (hashes) the value of the data at position p with the value of the data at the lagged offsets:\\
\verb"set(hash(buffer[p],buffer[p-offsets[k]],color));"\\
\indent
The fifth type of context makes pairwise combinations of the values of the data at each of the offsets:\\
\verb"set(hash(buffer[p-offsets[k1]],"\\
\verb"          buffer[p-offsets[k2]],color));"\\
\indent
The indices k1 and k2 run from 1 to n, with $k1<k2$.  There are $(n^2-n)/2$ of these.

The sixth type of context makes three-way combinations of the values of the data at each of the offsets:\\
\verb"set(hash(buffer[p-offsets[k1]],buffer[p-offsets[k2]],"\\
\verb"           buffer[p-offsets[k3]],color));"\\
\indent
The indices k1, k2, and k3 run from 1 to n, with $k1<k2<k3$.  There are $n(n-1)(n-2)/6$ of these.

In order to properly reconstruct the context model and decompress the data, the set of optimal contexts must also be stored.  This adds a few bytes to the representation, which can be incorporated into the file header.  In this implementation, a 4-byte long integer is stored for each element in the optimal context set.

\section{General Data Compression}
\verb" "
\indent
During the testing process it was verified that applying the definition of the autocorrelation function produces context models and results identical to those obtained by using the Weiner-Khinchin theorem, but the Weiner-Khinchin approach was many orders of magnitude faster, as expected.

The technique was tested on the image Rafale.bmp which is the benchmark image for lossless image compression at www.maximumcompression.com.  Without prior knowledge of the data format, PAQ8pxd v4 would compress the image to a size of 677,265 bytes (plus a 54-byte header, as determined by removing the custom image model) by using its default model for general data.  Using only the arguments of the set $A_{10}$ of the 10 largest values of the autocorrelation function (specifically, all 10 offsets were used for were used for context types 1-4, above, 7 offsets were used for type five, and 5 offsets were used for type six) our automatically generated context model demonstrates significantly improved general data compression, producing a file size of 624,775 bytes (plus a 54-byte header).

Based on these results, it is likely that the compression of a wide variety of signals could benefit from the use of autocorrelated contexts rather than adjacent, sequential contexts.  We will now consider an example of specialized models for image compression using autocorrelated contexts.

\section{Image File Compression}
\verb" "
\indent
The PAQ8pxd software produces a result of 525,663 bytes for this image when using its custom image model, which has evolved over several generations of the software to make advantageous combinations of neighboring pixels.  The three largest values of the autocorrelation function were used to construct the set $A_3$ of autocorrelated contexts, using the six types of contexts described above and used to demonstrate general data compression.  These contexts were combined with the hand-coded image model to produce a hybrid model.  This improves the overall result, yielding a slightly improved code of 523,932 bytes.

It should be noted, however, that the only notable input to the PAQ image model is the width of the image.  Moreover, since this value generally has one of the highest autocorrelations, it is typically among the leading values included in the set $A_n$, so the autocorrelated models can replicate the custom image model \emph{a priori}.  By substituting autocorrelated lags in place of width, the image model may be generalized.

If the width-dependent contexts of the PAQ image model are duplicated $n$ times and the autocorrelated offsets $A_n$ are substituted for the width, we may automatically generate an image context model which takes advantage not only of the dimensions of the image, but also of any other significant correlations which may be present.  This implementation uses the PAQ image model, which is conditional upon pixels directly to the left and above the current pixel, but also substitutes autocorrelated pixels into $n$ copies of these contexts, unless the autocorrelated pixels in question duplicate the contexts directly to the left of or above the current pixel, in which case they are ignored due to redundancy.  The significant differences in the contexts used in PAQ8pxd v4 and this implementation are the inclusion of the \verb"color" identifier and the omission of a counter in each autocorrelated context, as numerical experiments revealed that including the color was productive but that including the counter was counterproductive (the counter is retained in the original model, however, rather than the color).  For reference, the $8n$ autocorrelated contexts added to the model are, in the notation developed above, and with the C-style right-bitshift operator "\verb">>"":\\
\verb"set(hash(buffer[p-offsets[k]],color));"\\
\verb"set(hash(buffer[p-offsets[k]],buffer[p-1],color));"\\
\verb"set(hash(buffer[p-offsets[k]],buffer[p-1],buffer[p-2],color));"\\
\verb"set(hash((buffer[p-3]+buffer[p-offsets[k]]>>3,buffer[p-1]>>4, "\\
\verb"          buffer[p-2]>>4,color));"\\
\verb"set(hash(buffer[p-offsets[k]],buffer[p-1]-buffer[p-offsets[k]-1],"\\
\verb"          color));"\\
\verb"set(hash(buffer[p-offsets[k]]+buffer[p-1]-buffer[p-offsets[k]-1],"\\
\verb"          color));"\\
\verb"set(hash(buffer[p-(offsets[k]*3-3)], buffer[p-(offsets[k]*3-6)],"\\
\verb"          color));"\\
\verb"set(hash(buffer[p-(offsets[k]*3+3)], buffer[p-(offsets[k]*3+6)],"\\
\verb"          color));"\\

The set $A_6$ of autocorrelated contexts was constructed, but two of these contexts were the pixels directly above or to the left, so, in actuality, only four additional autocorrelated pixels were used.  All six contexts were stored, so the size of the resulting representation is 520,351 bytes (plus a 54-byte file header) which is thought to be a new record for the lossless compression of this file.  If the two redundant pixels to the left and above (which, perhaps not surprisingly, have the first and second largest autocorrelations, respectively) were explicitly excluded, as they will be in a subsequent implementation, the file would be losslessly compressed to a size of 520,343 bytes (plus the 54-byte header) with no loss of generality.

\section{Conclusion}
\verb" "
\indent
The Weiner-Khinchin theorem allows fast calculation of the autocorrelation function, and the largest values of this function may be utilized to provide a context for data.  We have seen that a direct application of context models automatically generated from the autocorrelation can produces results competitive with custom human-created context models, and that augmenting the human-created models with automatically generated context models results in superior performance.  Though the examples considered above concern image files, the method is well-suited for other types of unstructured signal data.  Audio, in particular, can benefit from the automatic generation of contexts, as the autocorrelation function provides a reliable means of identifying the fundamental frequencies of sound, even in the presence of noise.  These fundamental frequencies, being highly correlated, are well suited for use in context models for audio compression, which will be the topic of a subsequent paper.

\section{Disclosures}
\verb" "
\indent
This research was funded entirely by the author, John Scoville, and the method described is pending patent protection in the United States.

\bibliographystyle{amsplain}

\begin{thebibliography}{1}

\bibitem{VC92}
N.G.~Van Campen, \emph{Stochastic processes in physics and chemistry},
  North-Holland, New York, 1992.

\bibitem{Jaynes}
E.T. Jaynes, \emph{Probability theory: The logic of science}, Cambridge
  University Press, 2003.

\bibitem{MM00}
M.~Mahoney, \emph{Fast text compression with neural networks}, In Proc. FLAIRS,
  Orlando FL. (2000).

\bibitem{BWC89}
I.H.~Witten T.~Bell and J.G. Cleary, \emph{Modeling for text compression}, ACM
  Computing Surveys \textbf{(21)4} (1989), 557--591.

\end{thebibliography}
\providecommand{\bysame}{\leavevmode\hbox to3em{\hrulefill}\thinspace}
\providecommand{\MR}{\relax\ifhmode\unskip\space\fi MR }
\providecommand{\MRhref}[2]{%
  \href{http://www.ams.org/mathscinet-getitem?mr=#1}{#2}
}
\providecommand{\href}[2]{#2}

\end{document}